\documentclass[usegraphicx,usenatbib,natbib,useAMS]{mn2e}

\usepackage[T1]{fontenc}
\usepackage{aecompl}

\usepackage{graphicx}
\usepackage{epstopdf}
\usepackage{times}     
\usepackage{aas_macros}
\usepackage{amsmath}
\usepackage{amssymb}   
\usepackage{bm}
\usepackage{multirow}
\usepackage{verbatim}

\def\lesssim{\mathrel{\hbox{\rlap{\hbox{\lower4pt\hbox{$\sim$}}}\hbox{$<$}}}}
\def\gtrsim{\mathrel{\hbox{\rlap{\hbox{\lower4pt\hbox{$\sim$}}}\hbox{$>$}}}}

\def\logoh{\log(\text{O/H})}

\newcommand{\kms}{\ensuremath{{\rm km\,s^{-1}}}}

\makeatletter
\providecommand*{\diff}%
        {\@ifnextchar^{\DIfF}{\DIfF^{}}}
\def\DIfF^#1{%
        \mathop{\mathrm{\mathstrut d}}%
                \nolimits^{#1}\gobblespace
}
\def\gobblespace{%
        \futurelet\diffarg\opspace}
\def\opspace{%
        \let\DiffSpace\!%
        \ifx\diffarg(%
                \let\DiffSpace\relax
        \else
                \ifx\diffarg\[%
                        \let\DiffSpace\relax
                \else
                        \ifx\diffarg\{%
                                \let\DiffSpace\relax
                        \fi\fi\fi\DiffSpace}

\DeclareRobustCommand{\ion}[2]{%
\relax\ifmmode
 \ifx\testbx\f@series
  {\mathbf{#1\,\mathsc{#2}}}\else
  {\mathrm{#1\,\mathsc{#2}}}\fi
 \else\textup{#1\,{\mdseries\textsc{#2}}}%
\fi}

\title[]{Merging galaxies produce outliers from the Fundamental Metallicity Relation}

\author[A. E. Gr\o nnow, K. Finlator and L. Christensen]
       {Asger E. Gr\o nnow\thanks{E-mail: agronnow@dark-cosmology.dk}$^1$, 
        Kristian Finlator$^1$ and Lise Christensen$^1$\\
        $^1$ Dark Cosmology Centre, Niels Bohr Institute, University
        of Copenhagen, Juliane Maries Vej 30, 2100 Copenhagen, Denmark\\
}
\date{Accepted 2015 May 28. Received 2015 May 28; in original form 2014 September 19}
      
\pubyear{2014}

\begin{document}	

\maketitle

\label{firstpage}

\begin{abstract}
From a large sample of $\approx 170,000$ local SDSS galaxies, we find that the Fundamental Metallicity Relation (FMR) has an overabundance of outliers, compared to what would be expected from a Gaussian distribution of residuals, with significantly lower metallicities than predicted from their stellar mass and star formation rate (SFR). This low-metallicity population has lower stellar masses, bimodial specific SFRs with enhanced star formation within the aperture and smaller half-light radii than the general sample, and is hence a physically distinct population. We show that they are consistent with being galaxies that are merging or have recently merged with a satellite galaxy. In this scenario, low-metallicity gas flows in from large radii, diluting the metallicity of star-forming regions and enhancing the specific SFR until the inflowing gas is processed and the metallicity has recovered. We introduce a simple model in which mergers with a mass ratio larger than a minimum dilute the central galaxy's metallicity by an amount that is proportional to the stellar mass ratio for a constant time, and show that it provides an excellent fit to the distribution of FMR residuals. We find the dilution time-scale to be $\tau=1.568_{-0.027}^{+0.029}$ Gyr, the average metallicity depression caused by a 1:1 merger to be $\alpha=0.2480_{-0.0020}^{+0.0017}$ dex and the minimum mass ratio merger that can be discerned from the intrinsic Gaussian scatter in the FMR to be $\xi_\text{min}=0.2030_{-0.0095}^{+0.0127}$ (these are statistical errors only). From this we derive that the average metallicity depression caused by a  merger with mass ratio between 1:5 and 1:1 is 0.114 dex.
\end{abstract}


\section{Introduction}
Understanding the way that the metallicities of galaxies depend on galactic properties and events is crucial in understanding galaxy evolution as metallicity is connected to important galactic processes such as inflows, outflows and star formation. A relation between galaxy mass and metallicity was found in \cite{lequeux79} and measured with high precision based on SDSS Data Release 2 as a tight relation between galaxy stellar mass and gas-phase metallicity by \cite{tremonti04}. This mass-metallicity relation (MZR) is thought to arise from galactic winds being more efficient at blowing metals out of lower mass galaxies owing to their shallower gravitational potential wells \citep{dekel86}. The MZR depends on environment \citep{cooper08} and redshift \citep{erb06,maiolino08,henry13}. In addition \cite{ellison08a} found that galaxies with high star formation rates (SFR) showed systematically lower metallicities than galaxies with similar masses but lower SFRs. This was studied in more detail in \cite{mannucci10} (hereafter M10) who binned stellar masses and metallicities by SFR and found that the MZR anticorrelated with the SFR bin for masses $M_* < 10^{10.5} M_\odot$. In light of this, M10 introduced the Fundamental Metallicity Relation (FMR) between stellar mass, SFR and gas-phase metallicity. They interpreted the SFR dependence as being due to continuous accretion of pristine gas from the intergalactic medium raising the SFR while diluting the metallicity. Concurrently with M10 \cite{lara-lopez10} also investigated the SFR dependence of the MZR and also found a relation between mass, SFR and metallicity. These studies of second-parameter dependences of the MZR all look at galaxy samples that are selected to differ with the parameter of interest whose MZRs can then be compared. A complementary way of examining further dependences of the MZR or FMR is to instead select galaxies with abnormally low or high metallicity for their masses (and SFR in the case of the FMR) and then check what the metallicity offset correlates with.

\cite{peeples09} analysed a sample of 42 metal-poor galaxy outliers from the MZR and found that all but two of those showed signs of interaction. More systematically, M10 produced a histogram of the residuals of the FMR, i.e. the difference between the measured metallicity and the metallicity predicted by the FMR for each galaxy. These residuals closely followed a Gaussian distribution with the exception of an extended wing of galaxies with lower nuclear metallicities than predicted. There it was speculated that the bulk of these low-metallicity galaxies were interacting, but this was not examined further.

That interactions tend to dilute nuclear metallicities has been found both in observations of close galaxy pairs \citep{kewley06,ellison08b,michel-dansac08,scudder12} and in simulations of mergers \citep{montuori10,rupke10,perez11,torrey12} with these authors finding systematic offsets from the MZR of up to a few tenths of a dex towards lower metallicities for merging galaxies. The standard explanation for this phenomenon is that the centre of the primary spiral galaxy experiences a period of strong inflow of gas from the outskirts of the galaxy. This inflow is driven by torques exerted by stars in bar instabilities created by the tidal interactions \citep{mihos96} and it will be metal deficient compared to the nuclear metallicity as spiral galaxies have radial abundance gradients with lower metallicities at larger radii \citep{zaritsky94,luck11}. The inflow also leads to an increase in star formation which causes the metallicity to eventually recover on a time-scale of a few Gyr \citep{montuori10}.

In this paper, we ask whether the population of star-forming galaxies whose metallicities are significantly below expectations based on the FMR can be readily interpreted within the context of a simple model for galaxy mergers that dilute the nuclear metallicity and boosts its star formation. In particular:\\
(1) Can the low-metallicity tail be successfully modelled as being due to mergers?\\
(2) Do the galaxies in the low-metallicity tail show complementary evidence of being in mergers?\\
(3) What does this imply about the impact of mergers on galaxies' gas reservoirs and the time-scale over which mergers have such an impact?

We will show that mergers readily account for the observed low-metallicity outliers. This enables us to estimate the time-scale and magnitude of merger-induced metallicity dilution in a novel way.

In \S \ref{sect:sample} we describe our sample selection. In \S \ref{sect:data} we fit an FMR to our sample and examine the differences between the main sample and the tail. In \S \ref{sect:model} we review our merger model and in \S \ref{sect:modelfits} we describe how we find the best-fitting parameters and their values. In \S \ref{sect:discussion} we discuss uncertainties and assumptions in our model and compare our results to observations and hydrodynamical simulations. Finally, we summarize our findings and avenues for future work in \S \ref{sect:conclusions}.


\begin{figure*}
\includegraphics[width=0.9\textwidth,clip]{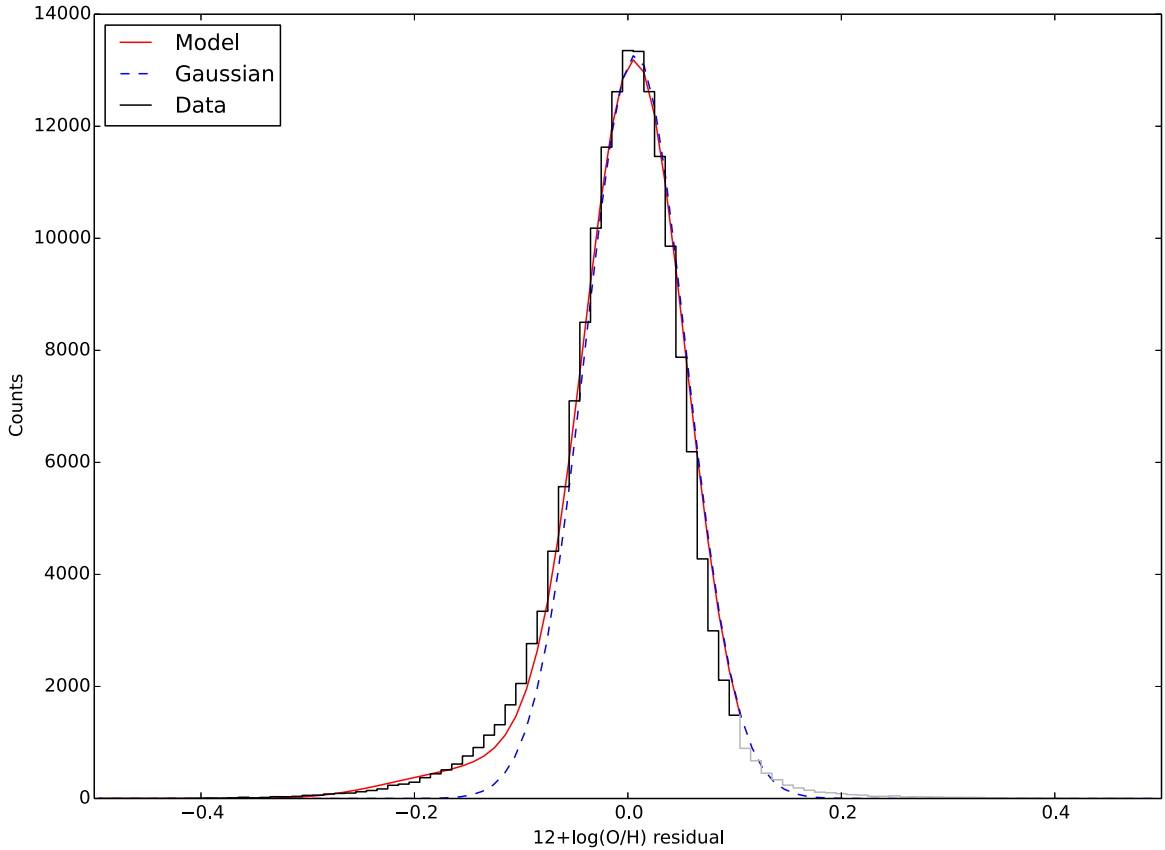}
\caption{The black histogram is the residuals of the fitted FMR, the unbroken line is the best-fitting model which has $\tau=1.57$ Gyr, $\alpha=0.248$ and $\xi_\text{min}=0.205$ and the dashed line is the best-fitting Gaussian. The grey part of the histogram at large positive offsets is the high-$r$ tail that is not included in the model.}
\label{fig:bestfitmodel}
\end{figure*}
\section{Sample selection}
\label{sect:sample}
We need to establish a large sample of galaxies in order to investigate the tail in the residuals of the FMR. We used a subset of the Sloan Digital Sky Survey (SDSS) DR9 \citep{ahn12} catalogue by the MPA-JHU group\footnote{'Max Plank institute for Astrophysics/John Hopkins University'. This group consisted of S. Charlot, G. Kauffmann, S. White, T. Heckman, C. Tremonti and J. Brinchmann.} available at http://www.sdss3.org/dr9/algorithms/galaxy\_mpa\_jhu.php where the techniques used to measure emission lines and derive galaxy parameters are also summarized. Half-light radii were adopted from the SDSS-DR9 photometric table ``PhotoObjAll''. The full catalogue contains 1,843,205 galaxies from which we selected 167,086 galaxies according to the criteria which were adopted from M10.

Only galaxies with redshifts within $0.07 < z < 0.30$ were selected to ensure that the 3 arcsec aperture of the spectroscopic fibre covered a significant part of the galaxies. We also demanded that H$\alpha$ was detected at a signal-to-noise ratio of at least 25. This ensures a sufficiently high S/N of the N[II] $\lambda$6584 line that is used in many metallicity calibrations as well as making BPT \citep{baldwin81} diagram classification of the galaxies more accurate. We selected only galaxies classified as BPT class 1 (star-forming) or 2 (low S/N star-forming) filtering out AGNs and composite galaxies. Finally, we filtered out galaxies for which it was not possible to measure the [OIII]$\lambda 5007$ line which is necessary to determine the metallicity using the O3N2 or R23 calibrations (see \S \ref{sect:FMRfit}); this amounted to 0.4 per cent of the remaining sample.	

Total stellar masses and SFRs were taken from the MPA-JHU catalogue. Stellar masses were calculated using the method of \cite{kauffmann03}, while the aperture-corrected SFRs were estimated through the method of \cite{brinchmann04} with slight modifications from \cite{salim07}. The masses were multiplied by 1.06 and the SFRs divided by 1.8 to scale them to a Chabrier IMF \citep{chabrier03}. Metallicities were measured within the aperture only.

\section{Data analysis}
\label{sect:data}
\subsection{Fitting an FMR}
\label{sect:FMRfit}
We had to use a strong-line method to find the metallicities of the galaxies in our sample. These methods are calibrated by fitting the relationship between the ratio of two or more strong emission lines and metallicities inferred directly from electron temperatures of HII regions. While ``direct'' metallicities found from electron temperature measurements are more accurate, they have to be calculated from auroral lines that are very weak and can only be detected with sufficient signal-to-noise in the SDSS spectra by stacking \citep{andrews13}. This is a problem as we need to be able to measure metallicities for single galaxies as we will be studying outliers. We chose to use the O3N2 calibration of \cite{marino13}, which gives $12+\logoh$ as a linear function of O3N2$\equiv \log\left([\text{OIII}]\lambda 5007/\text{H}\beta \times \text{H}\alpha/[\text{NII}]\lambda 6584 \right)$, to derive metallicities as this recent calibration is based on more extensive electron temperature data than older calibrations such as \cite{pettini04}. M10 used an average of the N2 and R23 calibrations of \cite{maiolino08} but as we are interested only in the residuals, the differences in the absolute metallicities derived from different calibrations do not matter and the differences in the residual distributions are quite small (see \S \ref{sect:calib}). We fitted an FMR of the form introduced in M10 to the data using least squares (see \S \ref{sect:parametrization} for a discussion of alternative parametrizations of the mass-SFR-metallicity relation). While M10 only fitted their FMR to galaxies with $M_*>10^{9.1} M_\odot$, we did not include a mass cut. In \cite{mannucci11}, where the FMR was extended down to $M_* \approx 10^{8.3} M_\odot$, it can be seen that the FMR for low-mass galaxies begins to deviate significantly from the extrapolation of the FMR of M10 at masses below $\sim 10^{8.8} M_\odot$. Only 0.1 per cent of the galaxies in our sample has such low masses. Our fit yielded
\begin{multline}
(12 + \logoh_\text{FMR}) = 8.504 + 0.169m - 0.034s - 0.110m^2\\ + 0.082ms - 0.048s^2,
\end{multline}
where $m \equiv \log(M_*)-10$ and $s \equiv \log($SFR$)$ and $m$ and $s$ are in units of $M_\odot$ and $M_\odot$ yr$^{-1}$, respectively. The standard errors of the coefficients are all between 0.1 and 1 percent except for the first coefficient which has a much smaller error, and the off-diagonal elements of the covariance matrix are much smaller than the diagonal elements. The residuals $r\equiv (12+\logoh_\text{data})-(12+\logoh_\text{FMR})$ were then computed for all galaxies.

We show a histogram of these metallicity residuals in Fig. \ref{fig:bestfitmodel}. This histogram has 100 equally spaced bins from $-0.5$  to $0.5$ dex. 58 of the 167,087 galaxies, i.e. about 1 in 3000 galaxies, fall outside this range. All but a single one of these outliers (which falls just below the range with $r =-0.5004$ dex) have $r > 0.5$ dex and appear to be mostly dwarfs with $\log(M_*/M_\odot)\sim 8$ and high specific SFRs (SSFRs). We will not examine these high-metallicity outliers further and from this point on in the analysis these outliers are filtered out.

This metallicity dispersion is almost Gaussian distributed but with a distinct tail towards lower metallicities as was also noted in M10 (cf. their fig. 3). We fit a Gaussian function to the metallicity dispersion using least squares finding a dispersion of $\sigma=0.048$ dex and a slight offset from zero of $\mu=0.007$ dex; this enables us to quantitatively define the tail as the bins with $r < \mu-2\sigma$. This tail contains an excess amount of 4.29 per cent of all galaxies relative to the number of galaxies that it would contain if it followed the fitted Gaussian. Upon close inspection a tail towards high metallicities can be seen as well but this tail only contains an excess of about 1 per cent of all galaxies (where this tail is defined as the bins with $r > \mu+2\sigma$).

M10 binned their stellar masses and SFRs in bins of 0.05 dex and then fitted their FMR to the median values in each bin. Doing this in our case turned out to make very little difference in the fitted FMR so we chose not to bin the masses and SFRs.

\begin{figure}
\includegraphics[width=0.49\textwidth,clip]{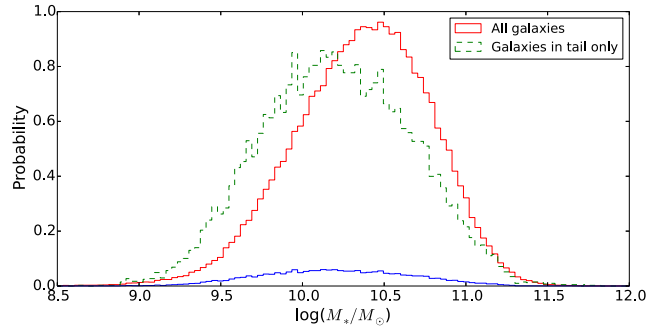}
\caption{Histograms showing the normalized mass distribution of the overall sample (red) and only the galaxies in the tail (green, dashed). The blue, low-probability curve is the mass distribution of the galaxies in the tail using the same normalization as for the overall sample to compare the sample sizes.}
\label{fig:massdist}
\end{figure}

\subsection{Properties of the galaxies in the tail}
We compare the galaxies in the tail to the general population to establish whether or not they form a distinct population. The stellar masses of the galaxies in the tail are generally lower than the general population with a median value that is 0.19 dex below the median value of all galaxies, both being approximately lognormally distributed (see Fig. \ref{fig:massdist}). This suggests that the tail cannot be an artefact of metallicity aperture effects alone (i.e. galaxies that are more completely covered by the aperture being found to have lower metallicities because of metallicity gradients) because in that case the mass distributions should be similar as the stellar masses are all estimated for entire galaxies. It also indicates that the tail is not predominantly driven by errors in mass because in that case the galaxies in the tail should have erroneously high masses as this would cause their predicted metallicities to be too high. That the galaxies in the tail have lower masses than the general population can be interpreted as being due to low-mass galaxies being more strongly impacted by mergers because of weaker bulges or higher gas fractions.

\begin{figure*}
\centering
\begin{minipage}[b]{.49\textwidth}
\includegraphics[width=0.99\textwidth,clip]{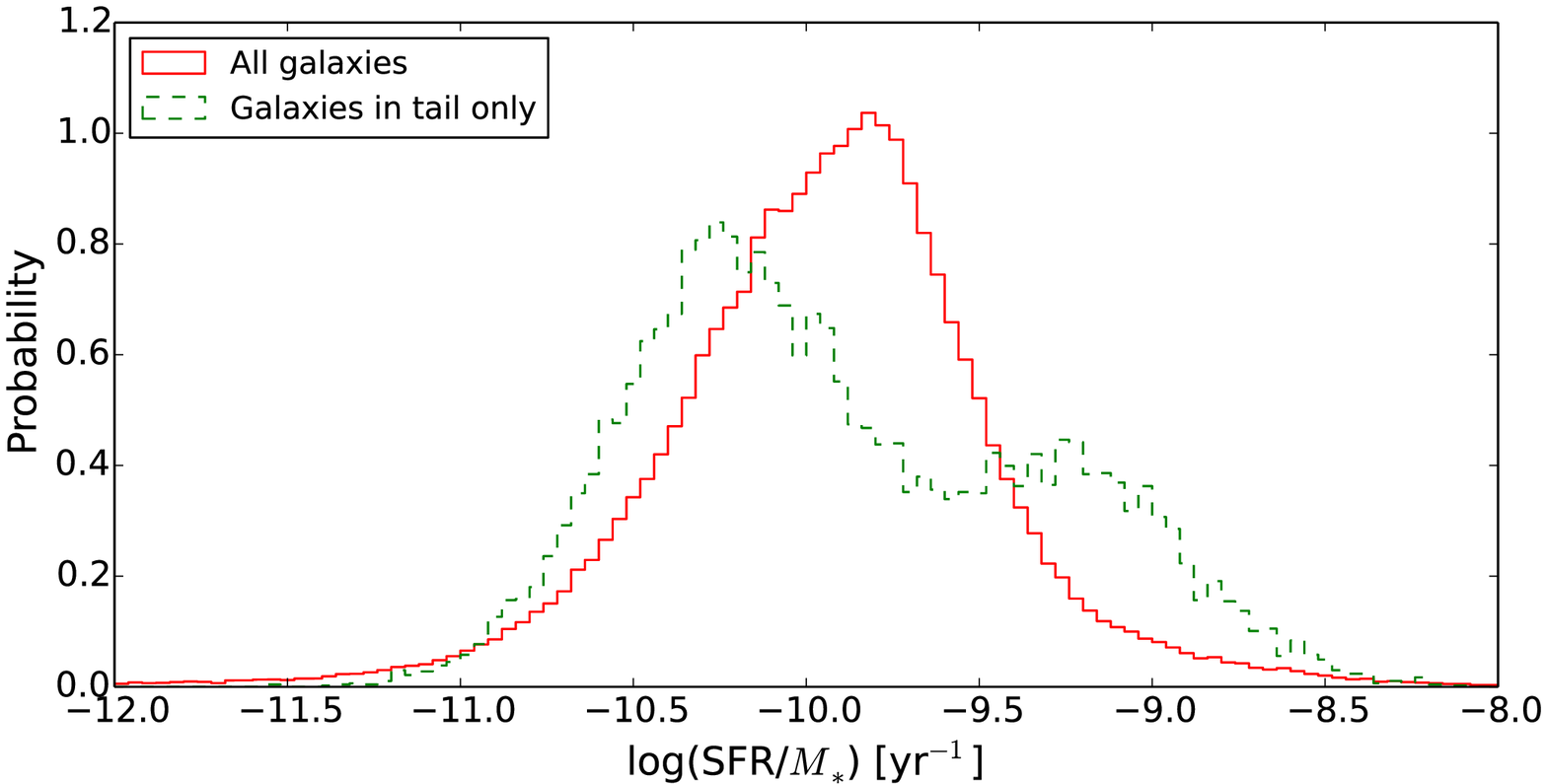}
\caption{Same as Fig. \ref{fig:massdist} but for the aperture-corrected SSFR (SFR/$M_*$).}
\label{fig:sfrdist}
\end{minipage}\quad
\setcounter{figure}{4}
\begin{minipage}[b]{.49\textwidth}
\includegraphics[width=0.99\textwidth,clip]{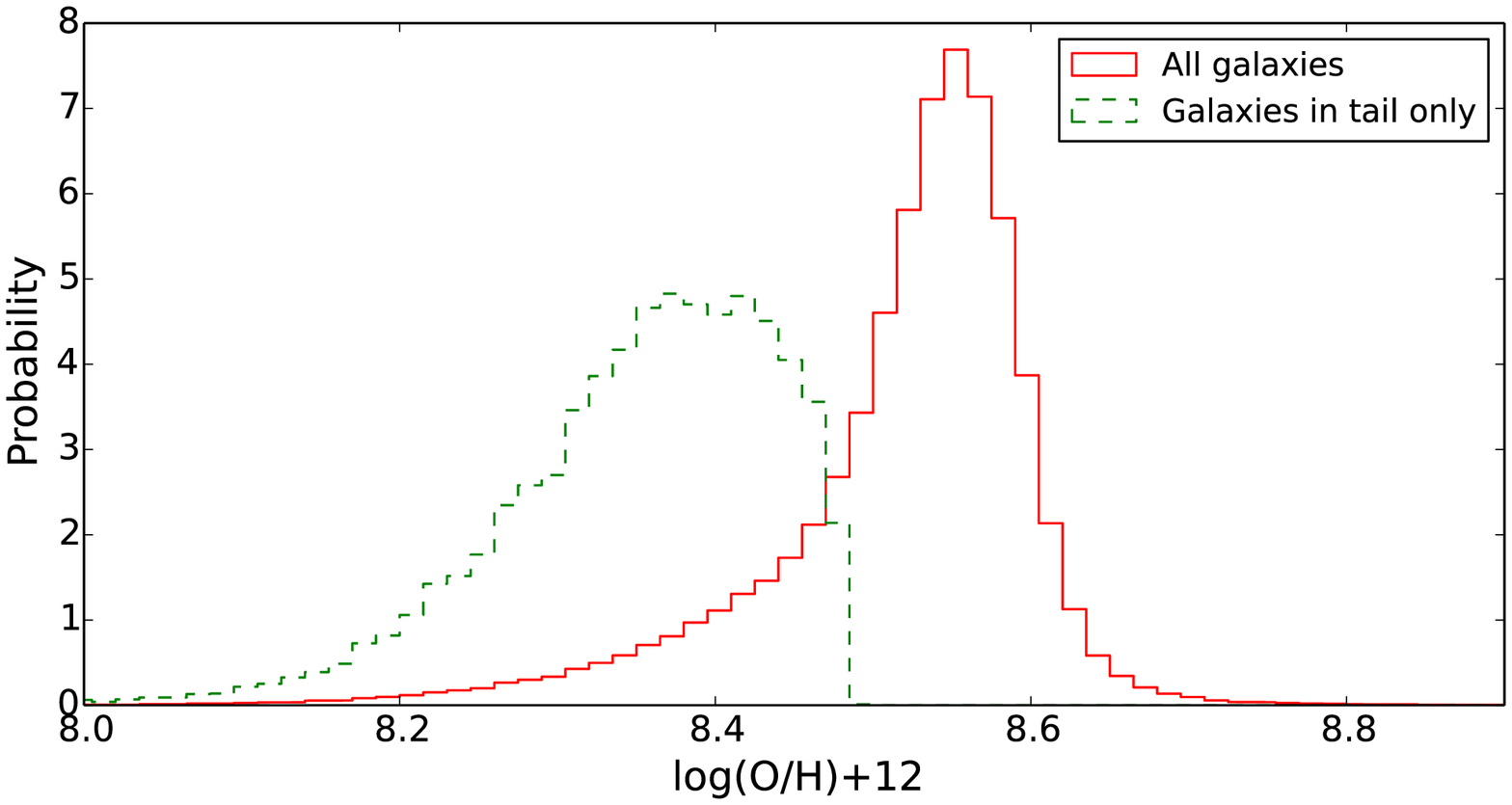}
\vspace{3pt}
\caption{Same as Fig. \ref{fig:massdist} but for metallicity.}
\vspace{6pt}
\label{fig:metaldist}
\end{minipage}\\
\setcounter{figure}{3}
\begin{minipage}[b]{.49\textwidth}
\includegraphics[width=0.99\textwidth,clip]{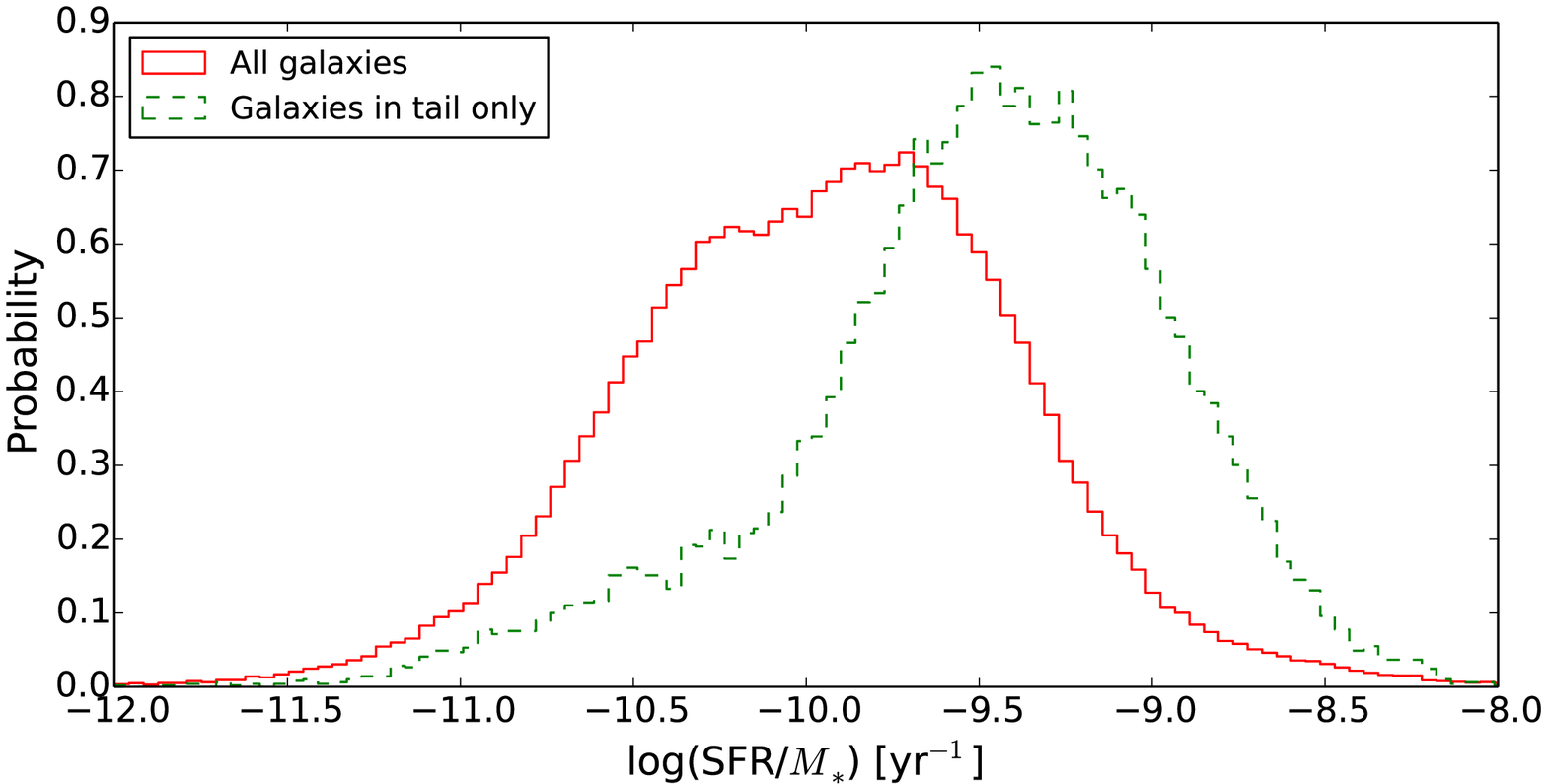}
\caption{Same as Fig. \ref{fig:massdist} but for the SSFR within the fibre.}
\label{fig:sfrdistfib}
\end{minipage}\quad
\setcounter{figure}{5}
\begin{minipage}[b]{.49\textwidth}
\includegraphics[width=0.99\textwidth, clip]{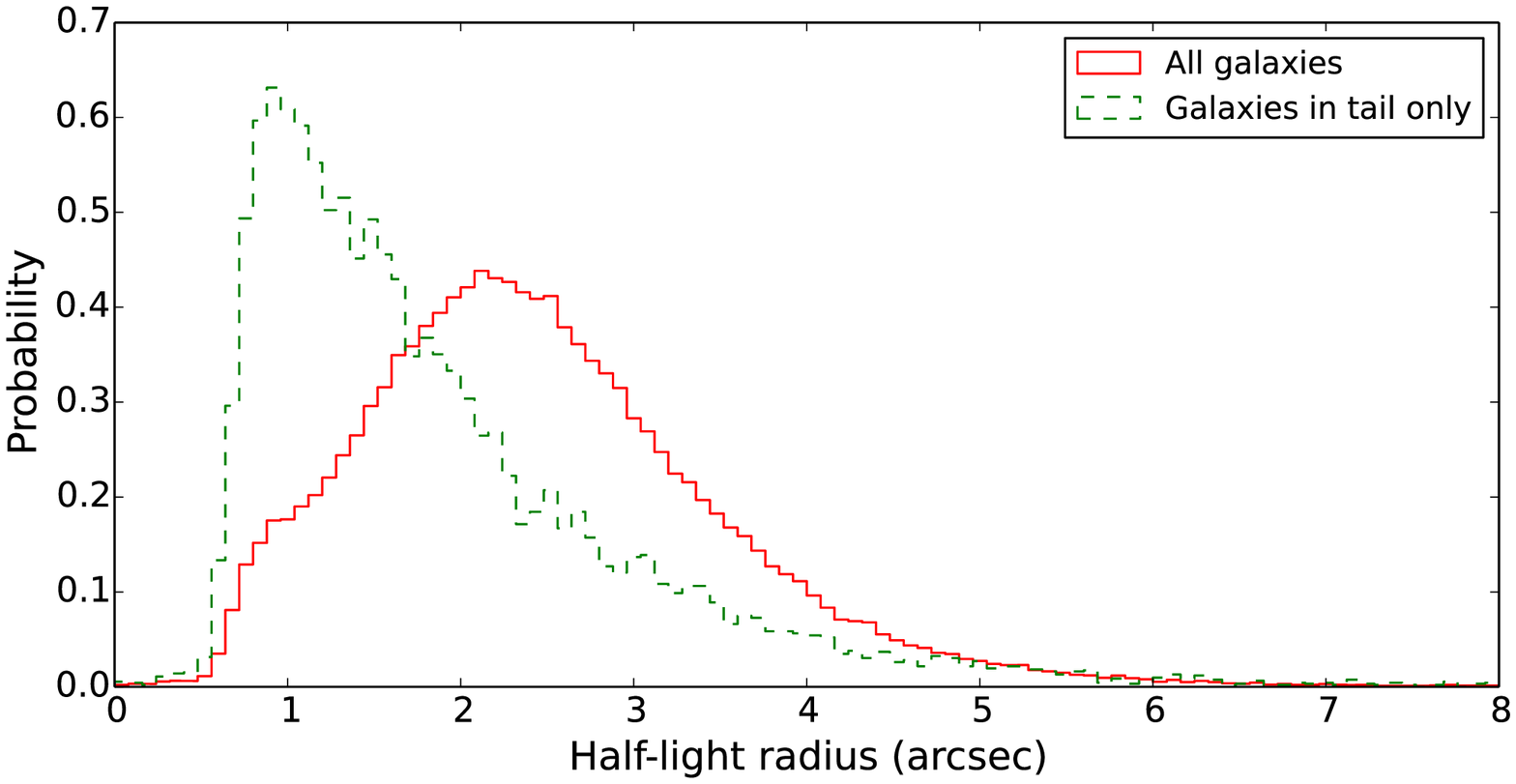}
\caption{Same as Fig. \ref{fig:massdist} but for U-band half-light radii.}
\label{fig:hlrdist}
\end{minipage}
\end{figure*}

The aperture-corrected SSFRs of the galaxies in the tail form a bimodial distribution with a peak about 0.5 dex below the mode for the general SSFR distribution and another, smaller peak about 0.6 dex above the mode for the general distribution (see Fig. \ref{fig:sfrdist}). Overall, the median value of the SSFRs of the galaxies in the tail is slightly below (0.08 dex) the median value for all galaxies. A simple interpretation of the high-SSFR peak in the tail is that many of those galaxies are experiencing a boost in star formation owing to interactions. In Fig. \ref{fig:sfrdistfib} we show the distribution of SSFRs within the aperture only. These were taken from the MPA-JHU catalogue as well and were derived using similar methods but without the photometric aperture correction. These are more accurate than the aperture-corrected SSFRs as the SFRs and masses are based only on spectroscopy. In this case, there is no bimodiality and the SSFRs of the galaxies in the tail are generally higher than the general population with a median value that is 0.5 dex above the median for all galaxies. Unlike the SSFRs derived in M10, where SFRs were estimated within an aperture while using global masses, these aperture SSFRs are physically meaningful. They show the relative amount of star formation in the same region as where the metallicity is measured \citep{salim14}. Enhanced SSFR (or, equivalently, enhanced SFR compared to control galaxies of similar mass) within the SDSS fibre has been linked to interactions \citep{li08,scudder12,patton13}. The origin of the bimodial shape of the aperture-corrected SSFRs in the tail is puzzling as none of the other parameters show any bimodiality and examining the subset of tail galaxies that are very close to one of the peaks reveals only slight differences. As there is also considerable overlap between the two SSFR peaks there is no simple way of filtering out either peak. A morphological study, that is outside the scope of this paper, would be necessary to check whether the low-SSFR tail galaxies are typically interacting.

The metallicities of the galaxies in the tail are as expected significantly lower than for the general population the median metallicity of the tail being 0.17 dex below the overall median value (see Fig. \ref{fig:metaldist}). One might worry that a significant part of the tail would then have metallicities below $12+\logoh=8.2$, which is the lower limit of validity for the O3N2 calibration we use. However, only 4.4 per cent of the galaxies in the tail have metallicities below this. For the overall sample 99.2 per cent are within the range of validity of $8.2 < 12+\logoh < 8.8$.

As can be seen in Fig. \ref{fig:hlrdist}, the galaxies in the tail have significantly smaller half-light radii than the general population. The figure shows the U-band half-light radii but the G and R bands show the same trend. The difference in the median values is about 30 per cent which is small enough that it could be attributed to the differences in the mass and redshift distributions of the tail and general population samples. However, the sudden rise at 0.5 arcsec suggests that the half-light radii of many of these galaxies are overestimated because they are smaller than the seeing. Additionally, as can be seen from the figure, the modes of the two half-light radius distributions differ by a factor above 2. Thus a significant part of the galaxies in the tail are significantly more compact than is typical for the full sample. This is in agreement with \cite{ellison08b} who found that galaxies in their pair sample with half-light radii below 3 kpc tended to have lower metallicities than galaxies with larger half-light radii.

In summary, galaxies whose metallicity is lower than predicted for their stellar mass and SFR have systematically low stellar mass, high SSFR within the aperture but slightly lower SSFR when aperture-corrected, small half-light radius and low metallicity. The tail towards lower metallicities is much larger than the tail towards higher metallicities. These considerations support the view that the low-metallicity tail consists of a physically distinct population.

\section{The merger model}
\label{sect:model}
We assume that the residuals of the FMR in the absence of mergers would be a normal distribution and fit a Gaussian function to the residuals from the FMR found in \S \ref{sect:FMRfit} using least squares. A Gaussian $G(r)$ is then fitted again but this time only in the interval [$\mu-2\sigma,\;\mu+2\sigma$], where the mean $\mu$ and the standard deviation $\sigma$ is estimated from the first fit, in order to avoid the low-$r$ tail where the shape deviates significantly from a Gaussian and the high-$r$ tail which is not a part of the model. We then introduce a simple model to take the effect of interaction-triggered metallicity dilution into account. This model has three free parameters, $\xi_\text{min}$, $\alpha$ and $\tau$, and two fixed parameters, namely the scatter and mean of the non-merging population, $G(r)$, that has already been fitted. $\xi_\text{min}$ represents the minimum mass ratio merger that can be discerned from non-merging population while $\alpha$ and $\tau$ represent the magnitude and time-scale of the dilution, respectively.

Mergers above a certain mass ratio threshold, $\xi_\text{min}$, shift galaxies towards lower $r$ by an amount $\alpha \xi_*$, where $\xi_*$ is the stellar mass ratio of the merger with respect to the most massive galaxy. So in the case of a 1:1 merger the metallicity changes by $-\alpha$ dex assuming that the change in $r$ caused by changes in stellar mass and/or SFR is small compared to the change in $r$ caused by the metallicity change (we show that this assumption is quite accurate in \S \ref{sect:discussmetal}). The metallicity change might of course vary with mass ratio in a more complicated way but as we have no knowledge of this relationship outside of that larger mass ratios should cause more dilution we assume this simple first-order relation. This offset remains constant for a time $\tau$ (in Gyr) before the galaxy's metallicity returns to normal, i.e. goes back to follow the Gaussian scatter around the FMR. $\xi_\text{min}$ can be thought of as representing the minimum shift in $r$ that can be distinguished from the intrinsic Gaussian scatter of the FMR, so we would expect $\alpha\xi_\text{min} \sim \sigma$. Note that our definition of the stellar mass ratio, $\xi_*=\frac{M_{*, \text{secondary}}}{M_{*, \text{primary}}}$, entails that $\xi_* \leq 1$, so secondary merger members (i.e. the less massive galaxy in the pair) are excluded from the model. This is intentional as we do not expect secondary merger members to experience notable metallicity dilution (see \S \ref{sect:mergerrate}).

The probability density function (PDF) of FMR residuals is given by the following expression:
\begin{multline}
\label{eqn:pdf}
P(r) = P(r \vert \text{unmerged}) P(\text{unmerged}) +\\P(r \vert \text{merged}) P(\text{merged}).
\end{multline}
Obviously, $P(\text{unmerged})=1-P(\text{merged})$. $P(\text{merged})$ is calculated by integrating the merger rate per galaxy per mass ratio per lookback time for a galaxy of stellar mass $M_*$, $\frac{\diff^2 P(\text{merged} \vert \xi_*, M_*)}{\diff \xi_* \diff t}$, over lookback times 0 to $T \gg \tau$, mass ratios from $\xi_\text{min}$ to 1 (as $\xi_*$ is defined as the stellar mass ratio of the merger with respect to the most massive galaxy) and the galaxy stellar mass PDF $\frac{\diff P}{\diff M_*}$. That is,
\begin{multline}
\label{eqn:Pmerged}
P(\text{merged}) =\\ \int_0^\infty \frac{\diff P}{\diff M_*} \int_{\xi_\text{min}}^1 \int_0^T \frac{\diff^2 P(\text{merged} \vert \xi_*, M_*)}{\diff \xi_* \diff t} \frac{\tau}{T} \diff t \diff \xi_* \diff M_*.
\end{multline}
$P(r \vert \text{unmerged})$ is just the normal distribution previously fitted to the residuals, $G(r)$, normalized to have unit area. 

We calculate the second term in $P(r)$ by integrating the probabilities given a specific mass ratio over $\xi_*$:
\begin{multline}
\label{eqn:Prmerged}
P(r \vert \text{merged}) P(\text{merged}) =\\ \int_{\xi_\text{min}}^1 P(r \vert \text{merged}, \xi_*) P(\text{merged}, \xi_*) \diff \xi_*.
\end{multline}
$P(r \vert \text{merged}, \xi_*)$ is just the Gaussian $P(r \vert \text{unmerged})$ where the mean is shifted from $\mu$ to $\mu-\alpha\xi_*$. As per equation \eqref{eqn:Pmerged}, equation \eqref{eqn:Prmerged} becomes
\begin{multline}
P(r \vert \text{merged}) P(\text{merged}) =\\ \int_0^\infty \frac{\diff P}{\diff M_*} \int_{\xi_\text{min}}^1 \int_0^T P(r \vert \text{merged}, \xi_*) \times\\ \frac{\diff^2 P(\text{merged} \vert \xi_*, M_*)}{\diff \xi_* \diff t} \frac{\tau}{T} \diff t \diff \xi_* \diff M_*.
\end{multline}

We obtain the merger rate per galaxy per stellar mass ratio per lookback time from the merger rate per halo per halo mass ratio per redshift (for a halo mass $M_h$) $\frac{\diff^2 P}{\diff \xi_h \diff z}(M_h, \xi_h, z)$ from \cite{fakhouri10} through the following calculation:

\begin{align}
\label{eqn:mergerrate}
\frac{\diff^2 P}{\diff \xi_* \diff t}(M_*, \xi_*, t) &= \frac{\diff^2 P}{\diff \xi_h \diff z}(M_h, \xi_h, z) \frac{\diff \xi_h}{\diff M_{*,s}}\frac{\diff M_{*,s}}{\diff \xi_*}\frac{\diff z}{\diff t}\notag\\
&= \frac{\diff^2 P}{\diff \xi_h \diff z}(M_h, \xi_h, t(z)) \frac{\diff M_h(M_{*,s})}{\diff M_{*,s}}\bigg|_{M_{*,s}}\times\notag\\
&\frac{M_*}{M_h(M_*)} (1+z)H_0\sqrt{\Omega_M(1+z)^3+\Omega_\Lambda},
\end{align}
where $M_{*,s}$ is the stellar mass of the smaller galaxy in the merger and $H_0$ is in units of Gyr$^{-1}$. We use a $\Lambda$CDM cosmology with $H_0=70\, \kms\, \text{Mpc}^{-1}$, $\Omega_\Lambda=0.7$ and $\Omega_\text{m}=0.3$ and the stellar mass -- halo mass relation of \cite{behroozi10}. Normally some ``merger delay'' time-scale would be imposed because a halo will spend some time as a subhalo before the merger is completed. We do not include this effect as the metallicity depression may well begin and end at different times than the merger itself as estimated from e.g. dynamical friction and operate on a time-scale that scales differently with parameters such as mass ratio from the merger delay time-scale (see \S \ref{sect:mergerrate}).

We evaluate equation \eqref{eqn:pdf} numerically creating a histogram of galaxies in bins of width $0.005$ dex in $r$. We do the time integration by simply multiplying by $\tau$ under the assumption that the merger rate is approximately constant over $T \gg \tau$ (i.e. assuming $T \ll t_H$ where $t_H$ is the Hubble time) with the lookback time set to $t(z=0.1)=1.30$ Gyr, the typical redshift of the galaxies in our sample. The stellar mass integration is done by looping over 10 stellar masses equally logarithmically spaced from $\log(M_*/M_\odot)=9.125$ to $\log(M_*/M_\odot)=11.375$ with $\frac{\diff P}{\diff M_*}$ being based on the mass distribution of our sample. Finally, the probability is converted to galaxy counts in each bin in order to compare the model to the data.

\section{Model fits}
\label{sect:modelfits}
We fit our five-parameter model in two steps: First, we fit a Gaussian to the FMR residual distribution in the way described in \S \ref{sect:model}. We found that the best-fitting Gaussian had a mean of 0.006 dex and a scatter of 0.046 dex. Then we fitted the three parameters describing the low-$r$ tail: $\xi_\text{min}$ which represents the minimum mass ratio merger that can be discerned, $\alpha$ which represents the metallicity change caused by a merger and $\tau$ which represents the time-scale of the dilution (see \S \ref{sect:model} for a detailed description of these parameters). We ran $40^3$ models with 40 equally spaced values of each of these.

We evaluate the relative goodness-of-fit of each model by defining a likelihood that compares the predicted and observed number of galaxies in each bin of $r$. In particular, we calculate each model's negative log-likelihood assuming that the errors are Poisson distributed. This is a more appropriate goodness-of-fit indicator than $\chi^2$ in this case because several of the bins in the low-$r$ tail have fewer than five galaxies and therefore have errors that are distributed significantly differently from a Gaussian and rebinning to ensure that all bins had at least five galaxies produced a binning that was too coarse. The negative log-likelihood for Poisson distributed errors is given by
\begin{equation}
-\ln{L} = \sum_{i \in r\text{ bin no.}} \ln{(d_i!)} + m_i-d_i\ln{m_i}.
\end{equation}
Here $d_i$ is the number of galaxies in bin $i$ and $m_i$ is the number of galaxies in bin $i$ predicted by the model. The set of parameters $\tau$, $\alpha$ and $\xi_\text{min}$ that produces the model that fits the data the best are the ones that minimize $-\ln{L}$ as this is equivalent to maximizing the likelihood $L$ (we call this maximum likelihood $L_0$). As $\ln{(d_i!)}$ becomes enormous and is constant across all the models anyway we set this term to zero and find the model that minimizes this shifted log-likelihood, which will be the same as the one that minimizes $-\ln{L}$, instead. We subtract this minimum shifted log-likelihood from all the shifted log-likelihoods to get the relative log-likelihood $\Delta \ln{L}=-\ln{L}-(-\ln{L_0})=\ln{L_0}-\ln{L}$. This is proportional to a likelihood ratio test, but we will stick to calling it relative log-likelihood and labelling this $\Delta \ln{L}$ as likelihood ratio tests are usually associated with the comparison of nested models.

\begin{figure*}
\includegraphics[width=0.8\textwidth]{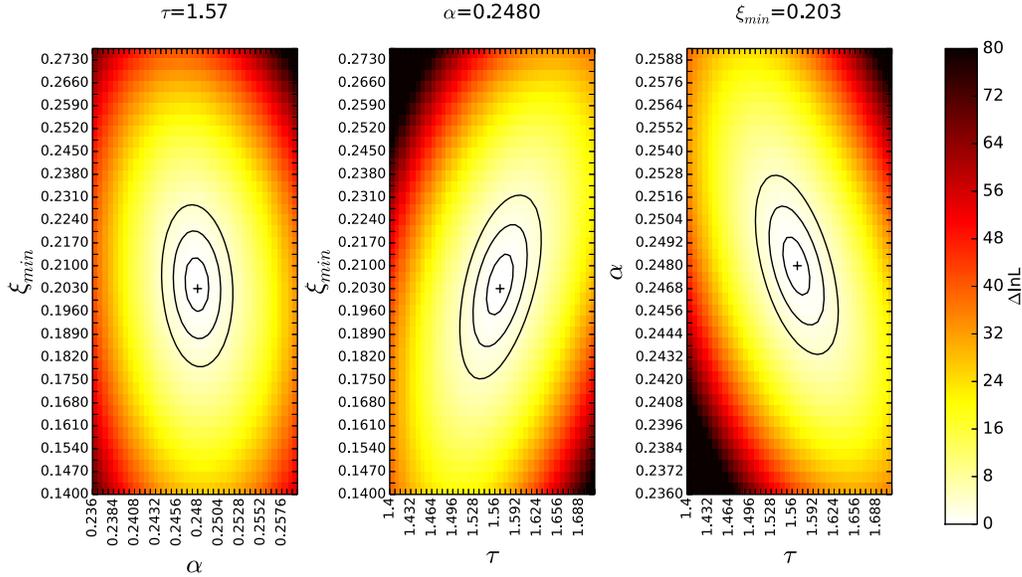}
\caption{Contour plots of the $\Delta \ln{L}$ parameter space with one parameter set to its best-fitting value. The crosses mark the minimum while the black overplotted contours mark 1-$\sigma$, 2-$\sigma$ and 3-$\sigma$ deviations from the minimum (i.e. $\Delta\ln{L}=0.5$, $\Delta\ln{L}=2$ and $\Delta\ln{L}=4.5$ respectively).}
\label{fig:lnLcontours}
\end{figure*}

\subsection{Parameters}
\label{sect:params}
Slices from the $\Delta \ln{L}$ space are plotted in Fig. \ref{fig:lnLcontours} for $\tau=[1.40,1.712]$, $\alpha=[0.2360,0.2594]$ and $\xi_\text{min}=[0.1400,0.2765]$ with 40 equally spaced values in each range, i.e. $\Delta\tau=0.008$ Gyr, $\Delta\alpha=0.0006$ and $\Delta\xi_\text{min}=0.0035$. These parameter ranges were found from a coarser initial fitting of models using the ranges $\tau=[0.5, 4.98]$, $\alpha=[0, 0.4985]$ and $\xi_\text{min}=[0.1, 0.88]$, also with 40 equally spaced values in each range. The best-fitting model has $\tau=1.568_{-0.027}^{+0.029}$ Gyr, $\alpha=0.2480_{-0.0020}^{+0.0017}$ and $\xi_\text{min}=0.2030_{-0.0095}^{+0.0127}$ and is plotted in Fig. \ref{fig:bestfitmodel}. The given errors are statistical errors only, the errors owing to various assumptions and uncertainties in the model are much greater (see \S \ref{sect:discussion}). As can be seen, there are slight degeneracies between $\tau$ and $\xi_\text{min}$ and between $\tau$ and $\alpha$; if $\tau$ is increased a higher $\xi_\text{min}$ and/or a lower $\alpha$ is preferred. This behaviour is quite intuitive: a higher number of galaxies showing diluted metallicities at a given time because the time-scale of metallicity depression is longer can be partially counteracted by excluding more merger mass ratios or making the effect of mergers on metallicity less severe.

The standard errors are found by marginalising the likelihood over the other two parameters for each parameter and finding the value of the parameter left and right of the peak where $L=e^{1/2} L_0$ ($L_0$ being the maximum likelihood) using linear interpolation between data points. This is done by numerically integrating the relative likelihood as found from $\Delta\ln{L}$ over the other two parameters. Using $\Delta\ln{L}$ rather than $-\ln{L}$ only changes the normalization of the marginalized likelihood distribution. For example, in the case of $\alpha$ the marginalized likelihood distribution is given by
\begin{align}
L(\alpha) &= \int_{\tau}\int_{\xi_\text{min}} \exp(\Delta \ln{L}(\alpha,\tau,\xi_\text{min})) \diff\tau \diff\xi_\text{min}\notag\\
&= \int_{\tau}\int_{\xi_\text{min}} \exp(\ln{L(\alpha,\tau,\xi_\text{min})}-\ln{L_0}) \diff\tau \diff\xi_\text{min}\notag\\
&= \frac{1}{L_0} \int_{\tau}\int_{\xi_\text{min}} L(\alpha,\tau,\xi_\text{min}) \diff\tau \diff\xi_\text{min}
\end{align}
$\frac{1}{L_0}$ can be found by requiring that the integral over all three parameters be unity.

\begin{figure*}
\includegraphics[width=0.75\textwidth]{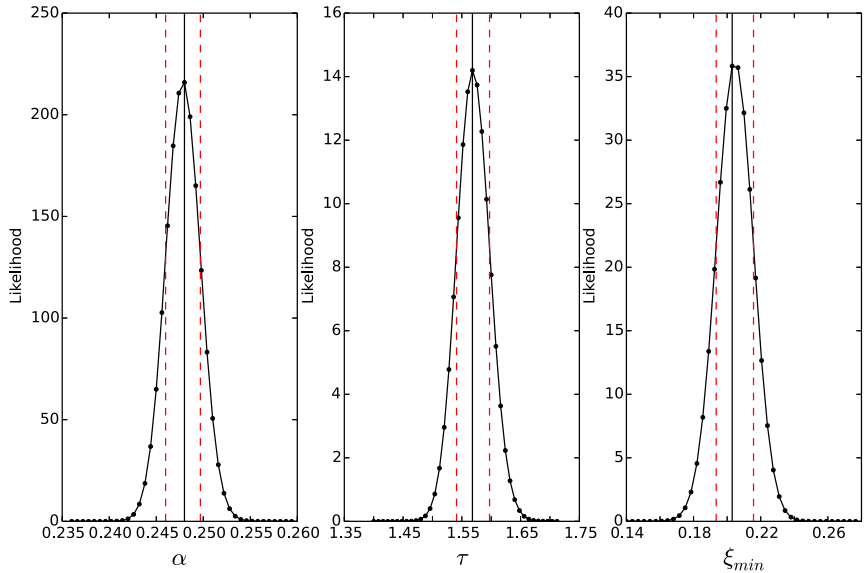}
\caption{Marginalized parameter space for $\alpha$, $\tau$ and $\xi_\text{min}$ from left to right. The unbroken line marks the parameter value that yields the maximum likelihood $L_0$ (with no interpolation between the parameter values used in the simulation). Dashed lines mark the formal standard errors as estimated by the linearly interpolated parameter values left and right of $L_0$ that lead to $L=e^{1/2}L_0$.}
\label{fig:params}
\end{figure*}

The normalized parameter likelihood distributions are shown in Fig. \ref{fig:params}. As can be seen, the distributions are approximately Gaussian and quite well resolved, meaning that the simple method used to derive the standard errors is appropriate, an assertion that is also supported by the fact that integrating from the left or right until 15.9 per cent of the area is enclosed yields nearly the same estimate of the errors.

\begin{figure*}
\includegraphics[width=0.96\textwidth,clip]{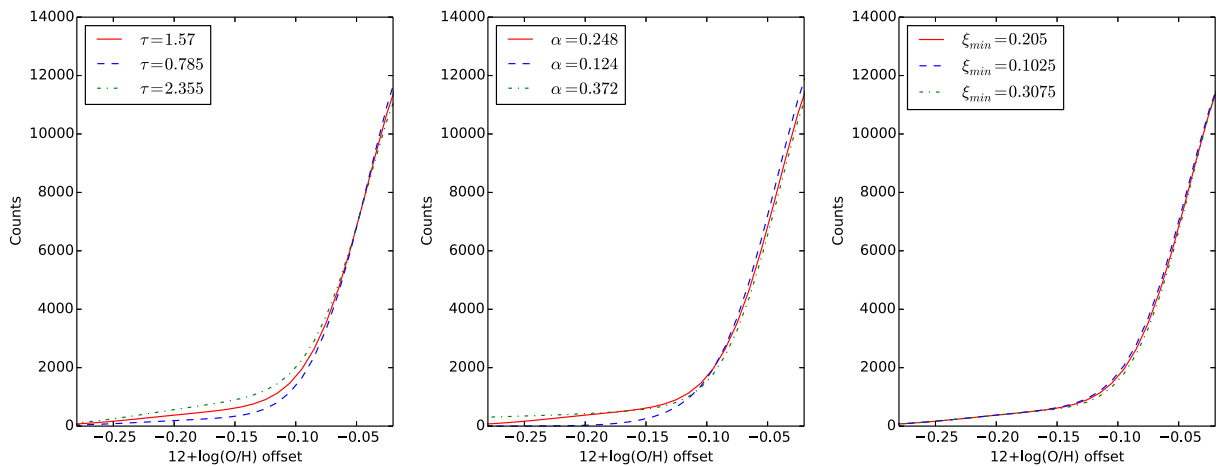}
\caption{The best-fitting model (unbroken, red line) has one of its parameters decreased by 50 per cent (dashed, blue line) and increased by 50 per cent (dot-dashed, green line) while the two other parameters and the normalization are held fixed. From left to right $\tau$, $\alpha$ and $\xi_\text{min}$ are varied.}
\label{fig:varyparams}
\end{figure*}

To get a feel for how each of the parameters affects the model, we change one of the parameters by $\pm$50 per cent while keeping the other two parameters at their best-fitting value and the normalization fixed. The resulting models are plotted in Fig. \ref{fig:varyparams}. As can be seen, increasing $\tau$ amplifies the tail while decreasing it brings the distribution closer to the best-fitting Gaussian. This is because increasing $\tau$ means that galaxies have diluted metallicities for a longer time and therefore the relative number of galaxies showing depressed metallicity at any one point in time increases. Varying $\alpha$ changes the overall shape of the tail with larger values causing a decrease of the moderate low-$r$ part and an increase of the more extreme offsets and smaller values bringing the distribution closer to a Gaussian that is wider than the best-fitting one. This is because increasing $\alpha$ increases the magnitude of metallicity dilution moving galaxies that are already experiencing dilution further towards lower metallicities. Increasing/decreasing $\xi_\text{min}$ diminishes/amplifies the moderate part of the tail while having no influence at the more extreme offsets. This is because increasing $\xi_\text{min}$ removes the galaxies at the lowest mass ratios previously included which therefore have the smallest metallicity dilutions.

\section{Discussion}
\label{sect:discussion}
\subsection{Magnitude and time-scale of the metallicity dilution}
\label{sect:discussmetal}
The magnitude of the nuclear metallicity dilution caused by a merger is connected to the parameter $\alpha$ in our model. Because the FMR depends on mass and SFR, the change in $r$ in our model is affected by changes in mass and SFR occurring during a merger, in addition to changes in nuclear metallicity. In practice, however, the dependence on mass and SFR of our fitted FMR is sufficiently weak compared to the differences in the tail relative to the general population that we can safely ignore this. The difference in median SFR between the general population and the tail by itself only leads to $\Delta r=0.004$ and the difference in mass only leads to $\Delta r=-0.026$. This is large compared to the statistical errors on $\alpha$ quoted in \S \ref{sect:params}, but insignificant compared to the systematic uncertainties discussed later in this chapter. Thus, we can equate the change in $r$ with the change in nuclear metallicity without introducing any significant uncertainty.

In addition, we look at the sample of 42 metal-poor galaxies of \cite{peeples09}. These galaxies were selected to be low-metallicity outliers from both the luminosity-metallicity and mass-metallicity relations and 40 of them were found to have disturbed morphologies indicative of interactions. 38 of these galaxies are in our sample and of those 30 of them are in the low-metallicity tail (with most of the rest being very close to the 2$\sigma$ cutoff). This shows that when taking the SFR dependence into account these interacting low-metallicity galaxies remain outliers.

The best-fitting value of $\alpha=0.248$ implies that the nuclear metallicity will on average decrease by about 0.25 dex during a 1:1 merger in agreement with the simulations of \cite{montuori10} and \cite{rupke10} who found metallicity depressions to be in the range 0.2 -- 0.3 dex and 0.1 -- 0.3 dex, respectively. Both of these were smoothed-particle hydrodynamics simulations of equal mass mergers with \cite{montuori10} being the more sophisticated of the two by including star formation and chemical enrichment from supernovae.

We calculate the average metallicity decrease for all mass ratio mergers included in a given model as $\int_{-0.5}^{0.5} r P(r \vert \text{merged})P(\text{merged})c_\text{norm} \diff r$ where $c_\text{norm}\equiv\left(\int_{-0.5}^{0.5} P(r \vert \text{merged})P(\text{merged}) \diff r\right)^{-1}$ (see \S \ref{sect:model} for the definitions of $P(r \vert \text{merged})$ and $P(\text{merged})$) and find that this is 0.114 dex for our best-fitting model. This is a somewhat more modest decrease than the 0.17 dex metallicity difference we find in the median metallicities between our general sample (which is very close to the difference between the averages) and the galaxies in the tail. However, this value is quite sensitive to the definition of where the tail begins, which we chose to be $2\sigma$ below the mean $r$ to be able to clearly distinguish the population in the tail from the general population. If, for example, the tail is defined to begin at $1.5\sigma$ below the mean $r$ instead, the galaxies in the tail have an average metallicity deficiency of 0.14 dex.

Comparing our average metallicity dilution of 0.114 dex to observations of close galaxy pairs in the SDSS, \cite{kewley06} found that pairs with separations $<20$ kpc followed a luminosity-metallicity (LZ) relation that was systematically shifted relative to the LZ relation for field galaxies by $-0.2$ dex but \cite{ellison08b} argued that roughly half of this shift is due to luminosity so correcting for this puts their observed metallicity dilutions close to our value. However, studies by \cite{ellison08b}, \cite{scudder12} and \cite{michel-dansac08} find a smaller change of $-0.03$ to $-0.05$ dex. Much of this discrepancy is due to the use of different lower mass ratio limits. \cite{ellison08b} and \cite{scudder12} both included mergers down to a mass ratio of 1:10 while our best-fitting model has $\xi_\text{min} = 0.205$, i.e. only includes mergers down to a mass ratio of about 1:5. Keeping the other parameters at their best-fitting value but setting $\xi_\text{min}=0.1$ yields an average metallicity depression of 0.087 dex. \cite{michel-dansac08} does not employ a sharp mass ratio cutoff but they do plot the MZR for all galaxies versus interacting galaxies for $\xi_* < 0.2$ and $\xi_* > 0.2$ with the larger interactions showing much stronger dilution (see their fig. 2).

The parameter $\tau$ should be the average time from the moment when the metallicity becomes diluted due to a merger until it completely recovers. Unfortunately most numerical simulations of merger-induced metallicity dilution are not run sufficiently long to estimate this time-scale and as such only yield lower limits. However we find our best-fitting value of 1.57 Gyr to be in good agreement with the typical metallicity depression time of $\sim 2$ Gyr found in \cite{montuori10} and above the lower limit of 1 Gyr found in \cite{torrey12}.

In our model, the metallicity depression is constant in time and lasts the same for all mergers. However, simulations show that for any particular merger both the magnitude and length of metallicity depression depends in complicated ways on the orientations of the galaxies and on whether they are on retrograde or prograde orbits. The metallicity depression time-scale probably also depends on mass ratio (see \S \ref{sect:aperture}). Furthermore \cite{torrey12} found that the initial gas fractions of the galaxies are important with higher gas fractions leading to less dilution. A characteristic feature found to some degree in all merger simulations is a ``double-dip'' shape of the metallicity as function of time associated with the first and second pericentric passages. We chose to ignore these complications as their effects are not well known and including any of them would add significant complexity to our model ruining its appealing simplicity. Furthermore, because we are using a rather large sample, our model should still yield representative average values of the dilution magnitude and time-scale.

\subsection{Merger rate}
\label{sect:mergerrate}
Galaxy-galaxy merger rates have significant uncertainties pertaining to the halo-halo merger rate, the stellar mass-halo mass relation used to convert halo masses to stellar masses and the method used to follow sub-haloes or the assumed merger delay, either of which are used to convert the halo-halo merger rate to a galaxy-galaxy merger rate (see \citealt{hopkins10} for a detailed analysis of each source of uncertainty in merger rates). For a halo-halo merger rate derived from a simulation such as the one we use from \cite{fakhouri10} (which is based on the Millennium and Millennium-II simulations) there are uncertainties from the definition of mass ratios, the construction of merger trees and the time resolution in the simulation. These yield a combined uncertainty of factor $\sim 2$.

A source of uncertainty that we introduce in converting this to a galaxy-galaxy merger rate is that we do not include any merger delay, i.e. we assume that the time that passes from one of the haloes to become a subhalo to the merger is completed is the same for all galaxies. This merger delay is often calculated based on a model of inspiral due to dynamical friction in which case it depends on mass ratio, virial radius of the primary galaxy and the energy and angular momentum of the orbit (see e.g. \cite{boylan-kolchin08} or \cite{jiang08}). Other methods based on characteristic time-scales for gravitational or angular momentum capture can also be used (see \cite{hopkins10b}). In general ignoring the merger delay time yields fewer major mergers. The differences in merger rates arising from using different merger delays or subhalo-following methods was examined in \cite{hopkins10} and \cite{hopkins10b} where it was found that the merger rate derived using no delay or method to follow subhaloes lay within the range of merger rates derived using these methods. More importantly we are interested in galaxies with diluted metallicities due to mergers which may well be a process that scales differently with various parameters and begins and ends at different times than the merger itself as measured from some dynamical friction or group capture time-scale.

At low stellar masses of $\log(M_*/M_\odot) \lesssim 10$, the gas fraction is significantly greater than at high stellar masses so our using the stellar mass ratio $\xi_*$ underestimates the actual metallicity impact, which should depend on the total tightly bound mass, of low-$M_*$ mergers by a factor of at least $\sim 3$ \citep{hopkins10}.

Qualitatively, the absence of red (i.e. passive) galaxies from our sample suppresses the merger fraction, as red galaxies have a higher merger rate at the median redshift of our sample than blue (i.e. star forming) galaxies \citep{lin08}. Thus, the actual merger rate should be less than what we obtain from equation \eqref{eqn:mergerrate} and consequently $\tau$ ought to be larger. However, the fraction of red galaxies in the relevant mass range should not be very large, particularly given that the galaxies in the tail tend to have lower masses than in general, so we expect this effect to be subdominant to the others evaluated in this chapter.

A consequence of defining mass ratio as $\xi_*=\frac{M_{*, \text{secondary}}}{M_{*, \text{primary}}} \leq 1$ is that we are ignoring the secondary members of mergers in our model. The degree of uncertainty caused by this depends on two factors: the number of such galaxies in our sample and how the metallicity of smaller companions in a merger changes. For a given stellar mass we can estimate the fraction of mergers as the primary galaxy from the merger rate and the galaxy stellar mass function. Based on this calculation and the stellar mass distribution of the galaxies in our sample we roughly estimate that if the secondary galaxy was affected similarly to the primary galaxy $\tau$ would be overestimated by a factor of $\sim$ 3 -- 4 in our model. This calculation is detailed in Appendix \ref{app:secondary}.

The effect on the metallicity of a galaxy caused by it merging with a more massive galaxy has not been examined in any detail but we would expect \emph{enrichment} rather than dilution in this case as the smaller galaxy accretes more enriched gas from the larger galaxy which also triggers star formation. Indeed \cite{scudder12} found that both members of merging pairs experienced similar levels of SFR enhancement and \cite{michel-dansac08} found that while massive galaxies in mergers showed diluted metallicities smaller merging galaxies with $\log(M_*)/M_\odot \lesssim 10$ showed enrichment instead. Therefore we expect the vast majority of galaxies that are merging or have recently merged as a secondary member to not be in the low-$r$ tail and the effect on our results of ignoring these should be small.

We can roughly estimate the merger fraction of our sample assuming that all galaxies in excess of the best-fitting Gaussian for the FMR residuals below the mean are merging. We find a merger fraction of 6.0 per cent of the sample in this way. As we are ignoring flybys (see \S \ref{sect:flybys}) and metallicities recover on a time-scale that is longer than the time-scale of the merger being visible (as either clear morphological disturbances or galaxy pairs) this is an upper limit. We cannot compare this directly to observational measurements of the merger fraction as these are measured for high masses (typically $M_* > 10^{10} M_\odot$) or luminosities only and for different types of samples (as we filtered out AGNs and galaxies without active star formation) at typically significantly higher redshifts. However, our value lies within the range of results in the literature found from morphological merger indicators (generally 1--10 per cent at lower redshifts; see \cite{lotz11}).

\subsection{Flybys}
\label{sect:flybys}
A potentially important complication that we are ignoring is flybys, i.e. when a galaxy passes another galaxy at a small distance but does not merge. We chose not to include flybys as the current quantitative knowledge of the effect and rate of flybys is poor and because of the considerable complexity it would add to our model. Little work has been done to examine the effects of flybys on metallicity. However \cite{montuori10} found that in their simulations of equal-mass interactions close flybys caused almost as much nuclear metallicity dilution as mergers and that this dilution lasted for almost as long as for mergers.

The flyby rate has also received very little study. \cite{sinha12} found from cosmological N-body simulations that the rate of ``grazing'' flybys (i.e. when two primary haloes approach each other, overlap for at least half a crossing time, then continue on different trajectories as two distinct primary haloes again) was comparable to the merger rate for halo masses $\log(M_h/M_\odot) \gtrsim 11$ (corresponding to $\log(M_*/M_\odot) \gtrsim 9$) at $z \lesssim 2$.

In the simple scenario where the flyby rate is equal to the merger rate and flybys dilute the nuclear metallicity in a way similar to mergers and on the same time-scale, the full effect of ignoring flybys in our model is simply that $\tau$ will be overestimated by a factor of 2. While this simple situation might not be that far from the truth considering the findings of \cite{montuori10} and \cite{sinha12} we expect the magnitude of dilution to depend on the pericentric distance of the flyby with greater distances causing less dilution.

\begin{table*}
\noindent\begin{tabular}{l l c c c}
\hline
  Calibration & Aperture-corrected SFR & $\tau$ (Gyr) & $\alpha$ (dex) & $\xi_\text{min}$ \\
  \vspace{3pt}
  \multirow{2}{*}{O3N2} & Yes & $1.57_{-0.027}^{+0.029}$ & $0.248_{-0.0020}^{+0.0017}$ & $0.205_{-0.0115}^{+0.0106}$ \\
  \vspace{3pt}
   & No & $1.79_{-0.042}^{+0.046}$ & $0.2095_{-0.0019}^{+0.0016}$ & $0.32_{-0.0135}^{+0.0149}$ \\
   \vspace{3pt}
  \multirow{2}{*}{R23} & Yes & $1.95_{-0.040}^{+0.038}$ & $0.360_{-0.0025}^{+0.0033}$ & $0.232_{-0.0098}^{+0.0144}$ \\
  \vspace{3pt}
   & No & $1.87_{-0.043}^{+0.035}$ & $0.346_{-0.0034}^{+0.0023}$ & $0.230_{-0.011}^{+0.014}$ \\
  \vspace{3pt}
  \multirow{2}{*}{N2} & Yes & $1.113_{-0.021}^{+0.023}$ & $0.2485_{-0.0019}^{+0.0022}$ & $0.245_{-0.0136}^{+0.0093}$ \\
  \vspace{3pt}
   & No & $1.04_{-0.034}^{+0.035}$ & $0.216_{-0.0025}^{+0.0021}$ & $0.382_{-0.0159}^{+0.0189}$ \\
  \vspace{3pt}
  \multirow{2}{*}{N2+R23} & Yes & $1.69_{-0.017}^{+0.026}$ & $0.2892_{-0.0020}^{+0.0017}$ & $0.127_{-0.0077}^{+0.0060}$ \\
  \vspace{3pt}
   & No & $1.04_{-0.034}^{+0.035}$ & $0.216_{-0.0025}^{+0.0021}$ & $0.382_{-0.0159}^{+0.0189}$ \\
   \hline
\end{tabular}
\caption{Best-fitting parameters for models based on data using different metallicity calibrations and with and without aperture correction of SFRs.}
\label{tab:systematics}
\end{table*}

\subsection{Metallicity calibrations}
\label{sect:calib}
\begin{figure*}
\includegraphics[width=0.9\textwidth]{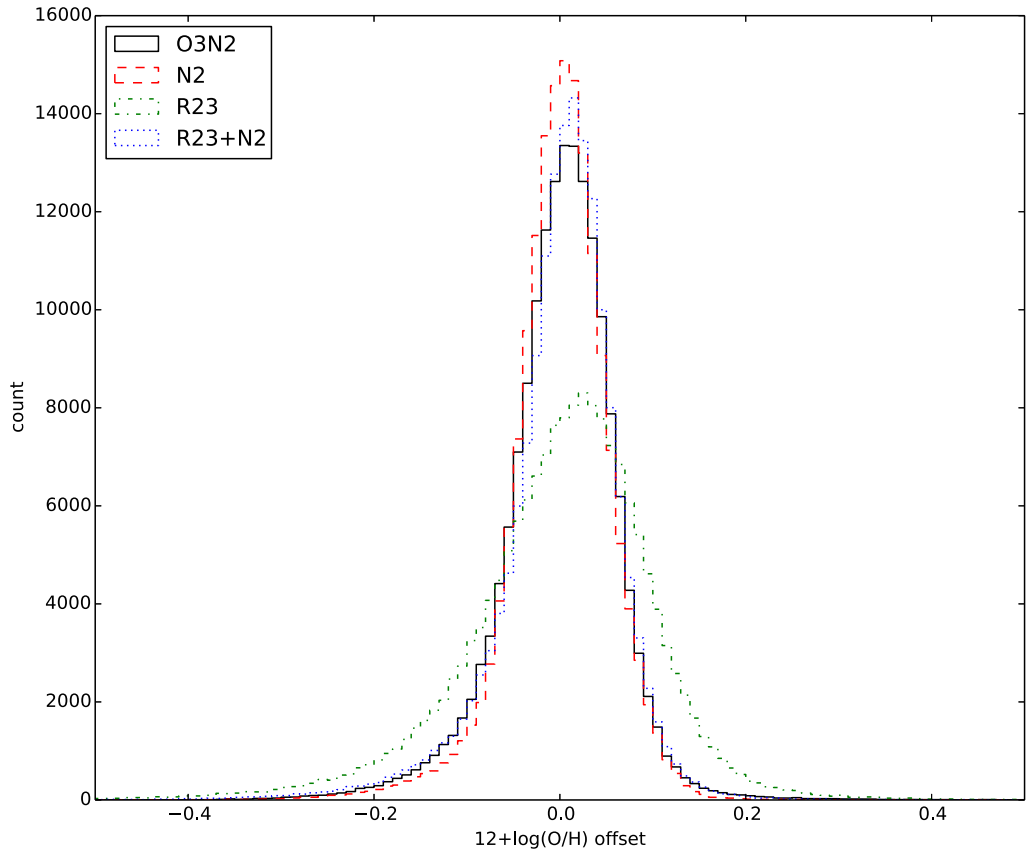}
\caption{The FMR residuals of different metallicity calibrations. The unbroken black line is the residuals of the O3N2 calibration of \protect\cite{marino13} that we have been using while the dashed red, dot-dashed green and dotted blue lines are residuals of calibrations based on N2 \protect\citep{denicolo02}, R23 \protect\citep{maiolino08} and the average of those two, respectively.}
\label{fig:calibrators}
\end{figure*}

As different metallicity calibrations yield quite different metallicities and MZRs (see \cite{kewley08}) we expect the choice of calibration to have an impact on the FMR. To evaluate the impact of different calibrations we compare the shape of the distribution of residuals of the fitted FMR and nature of the low-$r$ tail of the data based on the N2 calibration of \cite{denicolo02}, the R23 calibration of \cite{maiolino08} and the average of these two which was used in M10 with the previous data based on the O3N2 calibration of \cite{marino13}. We show the FMR residual distributions for these calibrations in Fig. \ref{fig:calibrators} and the best-fitting parameters of the corresponding models in Table \ref{tab:systematics}.

As with O3N2 these calibrations have limited ranges of validity, though both are broad. The range given for N2 in \cite{denicolo02} is $7.2 < 12+\logoh < 9.1$ and the entire sample lies within this. Recently, however, \cite{morales-luis14} have shown that N2 does not correlate with metallicity for galaxies with $12+\logoh < 7.6$. This is not an issue for our sample as only three galaxies have such low metallicities. The range for R23 is about $7.0 < 12+\logoh < 9.3$ and 99 per cent of the sample lies within this range.

\cite{denicolo02} find a relation for $12+\logoh$ as a linear function of N$2 = [\text{NII}]\lambda6584/\text{H}\alpha$. Using this relation to find the metallicities and fitting an FMR to these we find that the metallicity dispersion is not too dissimilar to the dispersion based on O3N2. Fitting a Gaussian distribution with least squares as before we find a slightly lower scatter and mean of $\sigma=0.042$ dex and a $\mu=0.004$ dex, respectively. The tail is smaller though, containing an excess of about 3.4 per cent of all galaxies. There is a systematic offset towards lower metallicities of 0.05 dex for the entire population and the tail alike. The SSFRs of the galaxies in the tail have the same bimodial distribution as for O3N2 but with a slightly lower median SSFR (0.03 dex lower than for O3N2, i.e. 0.11 dex lower than the median SSFR of the general population). As can be seen in Table \ref{tab:systematics}, the best-fitting value of $\alpha$ is almost the same as for O3N2 while $\tau$ is about 30 per cent smaller and $\xi_\text{min}$ is about 16 per cent larger representing a minimum mass ratio of about 1:4 rather than 1:5 for O3N2. There is a large overlap between which galaxies are in the tail for N2 and for O3N2: 58 (76) per cent of the galaxies in the O3N2 (N2) tail are also present in the N2 (O3N2) tail.

\cite{maiolino08} find a relation of $12+\logoh$ as a polynomial function of R$23 = ([\text{OII}]\lambda3727 + [\text{OIII}]\lambda4959 + [\text{OIII}]\lambda5007)/\text{H}\beta$. In this case we find a much wider and more skewed metallicity dispersion distribution with a scatter of $\sigma=0.80$ dex and mean $\mu=0.013$ dex. The tail is bigger containing an excess of about 6.8 per cent of all galaxies. As with N2 there is a systematic offset in the metallicities of 0.05 dex but towards higher, rather than lower, metallicities. The high-SSFR peak of the galaxies in the tail is less prominent than for O3N2 causing the median SSFR to be 0.08 dex lower (i.e. 0.16 dex lower than the median SSFR of the general population). The best-fitting values of the three model parameters are all larger than for O3N2, reflecting the wider tail, with $\alpha$ showing the biggest increase at 45 per cent. 70 (66) per cent of the galaxies in the O3N2 (R23) tail are also present in the R23 (O3N2) tail.

Finally, we look at the average of the N2 and R23 metallicities as was used in M10. In this case, the scatter is slightly smaller at $\sigma=0.045$ dex while the mean is slightly larger at $\mu=0.011$ dex. However, while the scatter is small the tail is large with an excess of 9.2 per cent of all galaxies. There is a systematic offset towards higher metallicities of 0.26 dex. The SSFRs in the tail have the same bimodial shape as for O3N2 with an almost identical median value as for O3N2 (0.01 dex lower than for O3N2, i.e. 0.09 dex lower than the general population). The best-fitting $\tau$ and $\alpha$ are a bit higher than for O3N2 while $\xi_\text{min}$ is significantly smaller representing a minimum mass ratio of about 1:8. Unlike the other calibrations for which the model can fit the residuals very well, in this case the fit is worsened by a few extreme outliers. 89 (66) per cent of the galaxies in the O3N2 (N2+R23) tail are also present in the N2+R23 (O3N2) tail. M10 used the N2 calibration of \cite{maiolino08} but we use the \cite{denicolo02} calibration because a sizeable fraction of the galaxies in our sample have $\log([\text{NII}]\lambda 6564/\text{OII}\lambda 3727) < 1.2$ for which the \cite{maiolino08} calibration is not valid. However, for the subset of galaxies that do have valid \cite{maiolino08} N2 metallicities the residual distribution is quite similar to the \cite{maiolino08} R23 metallicities.

From these comparisons, it is clear that while the shape of the metallicity dispersion around the FMR changes with the calibration used, a low-$r$ tail containing mostly the same low-metallicity galaxies is present across at least the three popular metallicity calibrations based on O3N2, N2 and R23. Thus, our qualitative conclusion that the tail is consistent with reflecting the impact of mergers is unchanged and the uncertainties in the inferred model parameters arising from different metallicity calibrations seem to be no greater than the uncertainties caused by e.g. uncertainties in the merger rate or ignoring flybys.

\subsection{Aperture effects}
\label{sect:aperture}
Spectroscopy is only available within the 3 arcsec aperture used for the SDSS fibre. Thus, we could only estimate metallicities for the central 3 arcsecs of the galaxies. The presence of metallicity gradients could therefore potentially lead to many of the metallicities being overestimated. To avoid significant aperture effects \cite{kewley05} recommended a minimum redshift cut of $z>0.04$ but \cite{kewley08} found this to be insufficient for galaxies with masses $M_* > 10^{10} M_\odot$. However, for our sample we use a significantly higher redshift limit of $z>0.07$. As the angular diameter distance at $z=0.07$ is about 70 per cent greater than at $z=0.04$, our lower redshift limit should be sufficient that aperture effects on the metallicity are relatively small for all masses.

The total SFRs used in this paper are the sum of the SFRs within the aperture, estimated using the method of \cite{brinchmann04} (these are based on several emission lines, but mainly on H$\alpha$), and the SFR outside the aperture, estimated from the photometry based on the assumption that the SFR has the same colour dependence outside of the aperture as within it. In contrast, M10 estimated the SFR only within the aperture from the H$\alpha$ line, as described in \cite{kennicutt98}. These SFRs are less uncertain than the total SFRs but only account for the star formation within a central diameter of 4-11 kpc of the galaxies in the redshift range used in M10 and this paper.

To examine the impact of using the aperture-corrected SFRs we fit FMRs using aperture-only SFRs, in addition to changing the metallicity calibrations, in table \ref{tab:systematics}. These SFRs are derived in the same way as in M10, correcting for dust using the Balmer decrement. The effect of aperture correcting the SFRs is typically to produce a larger but also more concave tail leading to higher best-fitting values of $\tau$ and $\alpha$ and lower values of $\xi_\text{min}$. As with different calibrations, the tail largely consists of the same galaxies when using the same calibration and switching between aperture-corrected and aperture-only SFRs, with 75--90 per cent overlap. Overall, the effect of aperture correcting the SFRs is typically smaller than the effect of changing the metallicity calibration.

\subsection{Parametrization of the FMR}
\label{sect:parametrization}
In this paper, we have followed M10 and expressed the relation between stellar mass, SFR and metallicity as a double quadratic. Usually the name ``fundamental metallicity relation'' refers specifically to this form. While this is a widely used parametrization, the question of whether it is the most appropriate form is unresolved.  Prominent alternatives include, for example, the ``Fundamental Plane'' \citep{lara-lopez10,lara-lopez13} and the relation based on the analytical model of \cite{lilly13}. The relation given by equation 40 of \cite{lilly13} closely follows the M10 relation for the sample used in M10 by construction; it is only outside the observed range of masses and SFRs that the \cite{lilly13} and M10 formulations differ significantly \citep{maier14}. As our sample resembles the M10 sample, differences with respect to the \cite{lilly13} relation will therefore be minimal.

In the Fundamental Plane introduced in \cite{lara-lopez10} the quadratic terms are dropped, leading to a linear relation between mass, SFR and metallicity. \cite{lara-lopez13} argues that the flattening in the M10 relation at high masses is an artefact caused by the line ratios used in their metallicity calibration saturating at high metallicities.  In this case, the underlying, physical relation could well be a simple plane. We do not attempt to evaluate whether the FMR is best described as a double quadratic or a plane (for a thorough comparison, see \cite{delosreyes15}). However, our merger model should also be able to fit the residuals reasonably well when a plane is used. Fitting a plane to our stellar masses, SFRs and metallicities using least squares, we find that the distribution of residuals has the same shape as for the M10 relation, i.e., the residuals follow a Gaussian distribution with an overabundance of galaxies with lower metallicities than predicted. The tail appears to consist of largely the same population of galaxies as when the M10 relation is used, i.e., the galaxies in the tail have slightly lower stellar masses, bimodial SFRs, lower metallicities and smaller half light radii than the general sample.

Finally, we have also fitted an MZR of the form given in \cite{tremonti04}, i.e. a fourth-order polynomial of the mass with no SFR dependence. In this case we also find that the residuals follow the shape of a Gaussian distribution with a wing towards low metallicities.

Generalizing the result that the tail could trace mergers to alternative functional dependences of metallicity on stellar mass, SFR and even further parameters is an avenue for future work.  This approach will be particularly useful once complementary insight is available from theoretical models that treat both small fluctuations in gas flows \citep{lilly13,forbes14} and galaxy interactions.

\section{Conclusions}
\label{sect:conclusions}
Fitting FMRs to our large sample of SDSS galaxies, we have shown that a tail in the distribution of the FMR residuals towards lower metallicities is present regardless of the metallicity calibration used. We develop a simple model where the tail consists of galaxies that are merging or have recently merged and show that this model is able to successfully reproduce the observed distribution of FMR residuals. The metallicity depressions and dilution time-scales that we derive by fitting this model are in good agreement with the results of merger simulations and observations of galaxy pairs. The success of this model suggests that the galaxies in the tail are recent mergers. This conclusion is supported by the fact that the galaxies in the tail form a distinct population with enhanced star formation within the aperture where metallicity is also measured.

We find that the average metallicity depression caused by a 1:1 merger is about 0.25 dex in agreement with the hydrodynamical simulations of \cite{montuori10} and \cite{rupke10}. We also find the average depression of all 1:5 to 1:1 mergers to be 0.114 dex which is consistent with the actual metallicity difference of the tail compared to the entire sample. This is a greater dilution than what is found in the metallicity measurements of SDSS pair samples of \cite{ellison08b}, \cite{scudder12} and \cite{michel-dansac08} but our value is consistent with those studies when the differences between the minimum mass ratios in those samples and in our model are taken into account.

We find that the average metallicity depression time-scale (the time from the onset of metallicity dilution until recovery to the pre-merger value) due to a merger is 1.57 Gyr in good agreement with the merger simulations of \cite{montuori10}.

Currently the quantitative knowledge of the rate and effect of galaxy flybys is very poor. However in future as our understanding of flybys increases, the precision of our model can be significantly improved by extending the formalism to include flybys in addition to mergers. In addition, future hydrodynamical merger simulations more comprehensive than the merger simulations done so far, that probe the entire range of galaxy masses and mass ratios could allow for a more specific estimation of the dilution time-scale. Provided such a simulation is run for a sufficiently long time so that any potential stellar mass and mass ratio dependence of the metallicity depression time-scale can be found it would allow our model to predict this time-scale for mergers of different member masses rather than as averaged over all mergers. In addition such a simulation might probe whether the mass ratio dependence of the metallicity depression (as time-averaged between pericentric passages) is approximately linear as we assume or has a more complicated form.

To further test if the galaxies in the tail are typically interacting, a morphological study of these galaxies could be undertaken, calculating merger indicators such as asymmetries or Gini coefficients of the luminosity distributions.

Due to the 3 arcsec aperture used in the SDSS, the accuracy of the galaxy parameters we use is limited by having to partly base stellar masses and SFRs on photometry, rather than spectroscopy only, and metallicities being estimated only within this aperture. In future, integral field spectroscopy should become available for a large fraction of our sample allowing us to obtain much more accurate global galaxy parameters. An important step in this direction is the MaNGA survey \citep{bundy15}, which is currently underway and will obtain integral field spectroscopy of a total of 10,000 nearby SDSS galaxies. While this only constitutes a relatively small fraction of our sample, it represents a factor $\sim 100$ increase in sample size compared to previous integral field spectroscopy surveys.

Our model makes some predictions that might be tested observationally. Low-metallicity outliers from the FMR should have a higher pair fraction than the general population and the mass ratios of pairs should be closer to unity as one moves to greater metallicity depressions. On the theoretical side, large-scale hydrodynamical simulations that track the metallicity of thousands of galaxies such as Illustris \citep{genel14,vogelsberger14} or EAGLE \citep{schaye15} should, in principle, enable a statistical study of merger induced metallicity dilution that could test whether the galaxies in the low-metallicity tail are really interacting (or have recently interacted) and in that case compare their inferred dilution time-scale, magnitude of metallicity depression and mass ratio dependence of those with our results. This is probably beyond the first generation of large-scale hydrodynamical simulations though, as outflows in Illustris are unphysically metal-depleted and EAGLE does not reproduce the MZR well \citep{schaye15}.

\section*{Acknowledgements}
The Dark Cosmology Centre is funded by the DNRF. We thank Sara Ellison and Filippo Mannucci for helpful discussions. We acknowledge the referee M. Peeples for useful suggestions which helped improve the paper.

Funding for SDSS-III has been provided by the Alfred P. Sloan Foundation, the Participating Institutions, the National Science Foundation, and the U.S. Department of Energy Office of Science. The SDSS-III web site is http://www.sdss3.org/.

SDSS-III is managed by the Astrophysical Research Consortium for the Participating Institutions of the SDSS-III Collaboration including the University of Arizona, the Brazilian Participation Group, Brookhaven National Laboratory, Carnegie Mellon University, University of Florida, the French Participation Group, the German Participation Group, Harvard University, the Instituto de Astrofisica de Canarias, the Michigan State/Notre Dame/JINA Participation Group, Johns Hopkins University, Lawrence Berkeley National Laboratory, Max Planck Institute for Astrophysics, Max Planck Institute for Extraterrestrial Physics, New Mexico State University, New York University, Ohio State University, Pennsylvania State University, University of Portsmouth, Princeton University, the Spanish Participation Group, University of Tokyo, University of Utah, Vanderbilt University, University of Virginia, University of Washington, and Yale University.

\bibliographystyle{apj}
\bibliography{lowzinteractions-publishedversion}

\appendix
\section{Estimating the potential impact of secondary merger members}
\label{app:secondary}
Given the merger rate per galaxy per stellar mass ratio per lookback time $ \frac{\diff^2 P}{\diff \xi_* \diff t}(M_*, \xi_*, t)$ for \emph{primary} merger members, i.e. $\xi_* \leq 1$, and the distribution of galaxy stellar masses $\phi(M_*)$ (defined such that $\phi(M) \diff M$ is the number of galaxies per volume with stellar mass between $M$ and $M+\diff M$) we can calculate the rate of mergers per galaxy that galaxies of a given mass experience as the \emph{secondary} member, i.e. $\xi_* > 1$.
\begin{multline}
\frac{\text{mergers}}{\text{time} \cdot \text{galaxy}}(\xi_*>1,M_0) =\\ \frac{\int_{M_0+\delta M}^\infty \phi(M_\text{pri}) \int_{(M_0+\delta M)/M_\text{pri}}^{M_0/M_\text{pri}} \frac{\diff^2 P}{\diff \xi_* \diff t} \diff \xi_* \diff M_\text{pri}}{\int_{M_0}^{M_0+\delta M} \phi(M) \diff M}
\end{multline}
where the merger rate is evaluated at $M_\text{pri}$, $\xi_*$, and $t=1.30$ Gyr which corresponds to $z=0.1$, the typical redshift of our sample. We use the mass function of \cite{baldry12}. If we assume that the metallicity of these galaxies is affected in the same way as the primary galaxies we can get a rough estimate of the overall error $\theta$ that ignoring secondary galaxies introduces in the dilution time-scale $\tau$. We define
\[R_\text{pri}(M) \equiv \frac{\text{mergers}}{\text{time} \cdot \text{galaxy}}(\xi_\text{min} < \xi_* < 1, M)\]
and
\[R_\text{sec}(M) \equiv \frac{\text{mergers}}{\text{time} \cdot \text{galaxy}}(\xi_*>1, M)\]
and calculate $\theta$ by integrating the fraction of merging galaxies that are primary at each mass over the mass distribution of our sample.
\begin{equation}
\theta = \frac{\tau_\text{corrected}}{\tau} = \frac{\int_0^\infty \frac{\diff P}{\diff M} \frac{R_\text{pri}}{R_\text{pri}+R_\text{sec}} \diff M}{\int_0^\infty \frac{\diff P}{\diff M} \diff M}
\end{equation}
For the best-fitting value of $\xi_\text{min}=0.205$ we find $\theta=0.29$ so $\tau$ would be overestimated by a factor $\sim 3-4$.

\end{document}